\documentclass[a4paper,11pt]{article}
\usepackage{jheppub}
\usepackage{subfigure}
\usepackage{graphicx}
\usepackage{subcaption}
\usepackage{caption}
\usepackage{cancel}
\usepackage{comment}

\usepackage{xcolor}
\usepackage[normalem]{ulem}

\newcommand{\ARdel}[1]{
    \ifmmode
        {\color{red}\cancel{#1}}
    \else
        {\color{red}\sout{#1}}
    \fi
}

\newcommand{\KGdel}[1]{
    \ifmmode
        {\color{blue}\cancel{#1}}
    \else
        {\color{blue}\sout{#1}}
    \fi
}

\newcommand{\BEdel}[1]{
    \ifmmode
        {\color{teal}\cancel{#1}}
    \else
        {\color{teal}\sout{#1}}
    \fi
}

\preprint{CERN-TH-2026-077}
\title{Asymptotic limits of constrained instantons}
\author{}

\abstract{We revisit the topic of false vacuum decay in field theory.  We focus on a toy model of a real massive scalar field with an unstable quartic potential.  This model has a false vacuum, and decay out of the false vacuum can be described via the method of constrained instantons, which work by introducing a constraint on the path integral.  We identify and develop three different asymptotic limits which enable analytic construction of approximate {constrained} solutions.  The first, in which the constrained solution is small compared to the inverse mass of the scalar field, is an application of the perturbative methods of Affleck, although we re-derive the main results and identify several terms which were previously neglected. Second, for very large constrained solutions we adapt the thin-wall approximation of Coleman.  However, we find that the large instanton limit does not always exist.  In this case we identify another useful limit, in which the Lagrange multiplier used to implement the constraint is large.  In this limit, the solution's scaling with the parameters may be found via dimensional analysis and an exact solution is obtained with a single numerical computation.
}

\author[a]{Benjamin Elder,}
\author[a]{Kinga Gawrych,}
\author[a,b,c]{and Arttu Rajantie}
\affiliation[a]{Abdus Salam Centre for Theoretical Physics, Imperial College London, London SW7 2AZ, United Kingdom}
\affiliation[b]{Theoretical Physics Department, CERN, 1211 Geneva 23, Switzerland}
\affiliation[c]{Rudolf Peierls Centre for Theoretical Physics, Oxford University, Oxford OX1 3PU, United Kingdom}
\emailAdd{b.elder@imperial.ac.uk}

\begin{document}

\maketitle

\section{Introduction}

Vacuum decay in field theory is normally described via instantons: classical ``bounce'' solutions to the Euclidean action~\cite{Coleman:1977py, Callan:1977pt}. These solutions allow one to compute the decay rate per unit volume per unit time out of a false vacuum in the saddle point approximation. However, it is possible to write a theory which clearly has false vacuum states which are unstable to quantum tunnelling, yet which admit no instanton solutions. One resolution is the method of constrained instantons \cite{Affleck:1980mp,Elder:2025kya}, in which one {generalises the standard saddle point method by} introducing an additional constraint on the {Euclidean} path integral. This procedure makes it possible to find instanton-like solutions, the contributions from which can be summed to find the false vacuum decay rate. The technical details of this method are discussed in \cite{Elder:2025kya}.

One simple example of a theory exhibiting this property is a real massive scalar $\phi^4$ theory with a negative quartic term \cite{Fubini:1976jm}. This theory has a false vacuum at $\phi = 0$, yet does not admit instanton solutions.\footnote{For proof see e.g.~\cite{Affleck:1980mp},\cite{Elder:2025kya}.} However, it is amenable to the method of constrained instantons.  In this method, one introduces a constraint 
in the path integral. Stationary points of this constrained path integral, which we refer to as ``constrained solutions'' can be found using the Lagrange multiplier method.
The vacuum decay rate of the original theory can then be obtained by integrating their contributions over the constraint.

In \cite{Elder:2025kya} we explored a numerical method for finding the constrained solutions and developed a formalism for determining whether a given solution contributes to the decay rate. However, 
in the limits of very small or large solutions, these numerical methods become increasingly costly, and on the other hand, simpler analytical approximations become available.

In this work we focus on analytic approaches to the same problem of constrained instantons. We do this for the same two constraint types as those explored in \cite{Elder:2025kya}, referred to as the cubic and hexic constraints. 
Our results rely on three distinct approximation regimes. First, both constraint types we consider admit the ``small-instanton'' limit described in~\cite{Affleck:1980mp}. We revisit the derivation of that work, in greater detail than presented originally. We identify several terms that were neglected previously, which ultimately account for a factor of 2 difference in the leading order correction compared to what appeared in~\cite{Affleck:1980mp}.  Comparison to our previous numerical results confirms the validity of this improved result. Second, the hexic constraint also admits a ``large-instanton'' limit, in which the transition takes place suddenly at a ``very large'' radius $r$. This strongly resembles the classic thin-wall approximation~\cite{Coleman:1977py} and is amenable to the same type of analysis. Third, we find that the cubic constraint does not admit the same ``large-instanton'' limit.  In fact, as the Lagrange multiplier $\kappa$ becomes arbitrarily large, instantons converge to a consistent size of $m \rho \approx 1$. We develop a new technique for this case, finding that in the large-$\kappa$ limit we can drop the quartic interaction term and extract the scaling of the action and constraint on $m$ and $\kappa$ within this limit. A single numerical solution for the field profile then provides exact results for all constrained solutions in that limit.

This paper is organised as follows.  In section~\ref{sec:constrained-instantons} we provide an overview of the method of constrained instantons and introduce the scalar field model and constraints we will use. In section~\ref{sec:small-instantons} we examine the ``small-instanton'' limit, which works for both types of constraint we consider. In section~\ref{sec:large-hexic} we explore the thin-wall approximation, which is applicable when there is a well-defined a ``large-instanton'' limit. In section~\ref{sec:large-cubic} we develop an approach for the case that there is no such limit, and instead identify the ``large-$\kappa$'' limit.
Finally, we make concluding remarks in section~\ref{sec:discussion}.

{\bf Conventions:} Throughout this work we use the 4D Euclidean action, therefore all indices are contracted with the all-plus Euclidean metric.

\section{The method of constrained instantons}
\label{sec:constrained-instantons}

We begin with a brief overview of the method of constrained instantons for describing vacuum decay.
The details of the method are discussed in \cite{Affleck:1980mp, Elder:2025kya}, but we reproduce the key steps here for the reader's convenience. We adopt as our toy model a real scalar field with a negative quartic potential which is unbounded from below. We include a positive mass term which creates a local minimum at $\phi = 0$,
\begin{equation}
    V(\phi) = \frac{1}{2} m^2 \phi^2 - \frac{\lambda}{4!} \phi^4~.
    \label{potential-massless-quartic}
\end{equation}
Clearly the field configuration $\phi(x)=0$ is classically stable but {at the quantum level becomes} unstable to large values of $\phi$ via tunnelling.  The configuration $\phi(x) = 0$ is thus a false vacuum.
To study tunnelling out of the false vacuum one normally looks for field configurations that are stationary points of the Euclidean action~\cite{Coleman:1977py}
\begin{equation}\label{eq:eucaction}
    S = \int d^4 x \left( \frac{1}{2} (\partial_\mu \phi)^2 + V(\phi) \right)~,
\end{equation}
that is, that are solutions to the Euclidean equations of motion.  

We are interested in the $4D$ spherically symmetric solutions, for which the equation of motion is
\begin{equation}
    \phi'' + \frac{3}{r} \phi' =  \frac{{\rm d} V(\phi)}{{\rm d} \phi}~,
    \label{instanton-eom}
\end{equation}
where a prime denotes a derivative with respect to the four-dimensional radial coordinate $r$.  In the ordinary picture of false vacuum decay, the tunnelling rate is described by the solutions to eq.~\eqref{instanton-eom} which satisfy the boundary conditions $\phi'(0) = \phi(\infty) = 0$.  The expected behaviour is that the field is at some finite value at the origin $\phi(0)$, and remains near that point until some value of $r$ at which the field rolls to a small value, approaching $0$ at asymptotically large values of $r$.  The point at which the field
rolls from large to small values is typically termed the ``size'' of the instanton.

In the special case of a massless theory, $m = 0$, the equation of motion~\eqref{instanton-eom} becomes
\begin{equation}
    \phi'' + \frac{3}{r} \phi' = - \frac{\lambda}{6} \phi^3~,
    \label{fubini-eom}
\end{equation}
and the solution is
\begin{equation}
\label{equ:fubini}
    \phi_0(r) = \frac{4 \sqrt 3 \rho \lambda^{-1/2}}{\rho^2 + r^2}~,
\end{equation}
This solution is referred to as the Fubini instanton~\cite{Fubini:1976jm}.\footnote{Sometimes it is called the ``scale-invariant instanton'' due to the classical scale invariance of the massless theory.} The parameter $\rho$ is referred to as the \emph{size} of the instanton, and can take any real value, which is a consequence of the scale invariance of the massless theory.
The Euclidean action of each instanton is independent of $\rho$,
\begin{equation}
    S_0 \equiv S[\phi_0] = \frac{16 \pi^2}{\lambda}~.
\end{equation}

If the mass term is finite then there is no solution to eq.~\eqref{instanton-eom}, that is, there is no stationary point of the Euclidean action eq.~\eqref{eq:eucaction}. A simple scaling argument demonstrates this: {given any field configuration $\phi(r)$, we can scale it via $\phi(r) \to a \phi(a r)$.} The action scales as
\begin{equation}
    S \to \int d^4 x \left( \frac{1}{2} (\partial \phi)^2 - \frac{\lambda}{4!} \phi^4 \right) + a^{-2} \int d^4 x \frac{1}{2} m^2 \phi^2 ~,
    \label{action-scaling}
\end{equation}
The first term is composed of scale invariant pieces and remains unchanged.  On the other hand, the second integral is always non-negative and only zero if $\phi = 0$, so any $a > 1$ reduces the action continuously.
Therefore the only stationary point of the Euclidean action is the trivial field configuration $\phi(r) = 0$.
This does not mean that there is no tunnelling out of the false vacuum, as one only needs to visualise the potential to see that $\phi = 0$ must be unstable to decay via quantum mechanical tunnelling.

However, the saddle point approximation is only one possible way to evaluate the path integral, and when it fails, one needs an alternative method.
The method of constrained instantons \cite{Affleck:1980mp} is
{one such alternative}.  In this method, an integral over a $\delta-$function is inserted in the path integral
in such a way that the resulting constrained path integrals have saddle points.
The method is described in \cite{Elder:2025kya}, although we recount the main points here.  The false vacuum decay rate $\Gamma$ is given by
\begin{equation}
    \Gamma = \frac{2}{\cal V} \mathrm{Im} \log Z~,
\end{equation}
where ${\cal V}$ is a spacetime volume and $Z$ is {the Euclidean} path integral
\begin{equation}
    Z = \int {\cal D} \phi e^{- S[\phi]}~,
\end{equation}
with the boundary condition $\phi\rightarrow 0$ at infinity.

We can define a constraint as
\begin{equation}
    \xi[\phi] = \int d^4x {\cal O}(\phi)~,
    \label{constraint-definition}
\end{equation}
where ${\cal O}$ is some operator that depends on the field $\phi$.
We can insert an integral over a $\delta-$function into the path integral, obtaining
\begin{equation}
    Z = \int {\rm d} \bar \xi  \int {\cal D} \phi~ \delta (\xi[\phi] - \bar \xi) e^{- S[\phi]}~.
\end{equation}
To perform the innermost integral, we restrict our attention to field configurations that satisfy the constraint $\xi[\phi] = \bar \xi$.  We can use the method of Lagrange multipliers to build a {\it modified action} that has the same stationary points as the innermost integral:
\begin{equation}
    \tilde S_\kappa = \int d^4x \left( \frac{1}{2} (\partial \phi)^2 + V(\phi) \right)
    + \kappa \int d^4x\, {\cal O}(\phi)~.
    \label{action-general-constraint}
\end{equation}
We have added the constraint term to the original action, along with a Lagrange multiplier $\kappa \in \mathbb{R}$.  As long as ${\cal O}$ is not scale invariant, and scales differently from the mass term under the scaling transformation of eq.~\eqref{action-scaling}, the modified action will have nontrivial saddle points.  This allows one to use the familiar machinery of instantons to describe vacuum decay, even in theories that do not admit bounce solutions.\footnote{We emphasize that constrained solutions are not instantons of the original theory.  An instanton is a saddle point of the original action, and constrained solutions are not.  Furthermore, an instanton has a single negative mode in its spectrum, while the same is not always true of constrained solutions and needs to be checked carefully~\cite{Elder:2025kya}.  This latter property is essential for a given field configuration to contribute to the path integral for tunnelling.}  Our main interest in this work is to find the saddle points in the constrained path integral, while the task of performing the rest of the path integral will be explored in upcoming work~\cite{prefactorPaper}.

For future reference, we will at times refer to the {\it modified potential}
\begin{equation}
    \tilde V_\kappa(\phi) = V(\phi) + \kappa {\cal O}(\phi)~.
    \label{modified-potential}
\end{equation}

The equation of motion for the modified action is
\begin{equation}
    \phi'' + \frac{3}{r} \phi' =  \frac{{\rm d} V(\phi)}{{\rm d} \phi}+\kappa \frac{{\rm d} \mathcal{O}(\phi)}{{\rm d} \phi}~,
    \label{eom-constrained}
\end{equation}
One can then solve the equation of motion for any $\kappa$, which corresponds to the range of possible constraints $\bar\xi$.  Finally, the tunnelling rate may be obtained by integrating over all constrained solutions with valid constraints $\bar\xi$ in the path integral~\cite{Elder:2025kya}.

\begin{figure}
    \centering
    \includegraphics[width=0.75\linewidth]{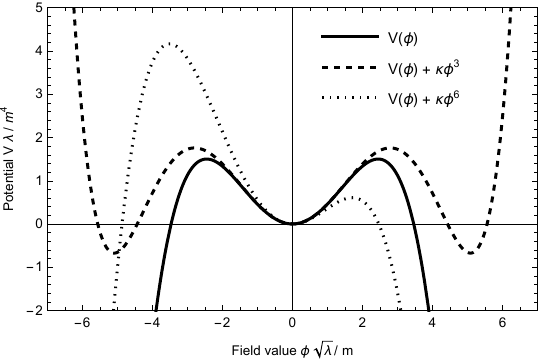}
    \caption{Diagram of the scalar potential $V(\phi)$ and the modified potentials we consider in this work.  An arbitrary value of $\kappa$ was chosen in order to represent the shape of the modified potentials.}
    \label{fig:potentials}
\end{figure}

We will focus on monomial constraints ${\cal O}(\phi) = \phi^n$.  To study a range of possible behaviour, we specialise to two choices of $n = 3, 6$:
\begin{align} \nonumber
    {\cal O}_\mathrm{cubic}(\phi) &= \phi^3~, \\
    {\cal O}_\mathrm{hexic}(\phi) &= \phi^6~.
\end{align}
These will be referred to as cubic and hexic constraints, respectively.  The potential, and the modified potentials, are sketched for a representative value of $\kappa$ in figure~\ref{fig:potentials}.
Each of these is explored for the entire allowed range of $\kappa$ and corresponding $\bar\xi$ values in the remaining sections.

\section{Small instantons: expanding in the massless limit}
\label{sec:small-instantons}

The limit in which the instanton size is small compared to the Compton wavelength $m^{-1}$ was explored in detail in~\cite{Affleck:1980mp}.  Here we recall some of the salient points, and apply them to the constraints under consideration in this work.
In the process, we identify several terms that were previously neglected in~\cite{Affleck:1980mp}, ultimately leading to a factor of $2$ difference in the first correction to the action.

The essential idea is the following.  We assume that the instanton size $\rho$ is well-defined, and that $m \rho \ll 1$.  This means that the field rolls to a small value at a radius much shorter than the Compton wavelength.  Before the field rolls, the potential is dominated by the quartic term, so the field profile should strongly resemble the Fubini instanton at small radii.  After the field rolls, the field value approaches $0$ and the potential is dominated by the mass term.  The Fubini instanton is thus a good approximation at small radii, with corrections that are suppressed by $m \rho$, and ceases to be a good approximation at larger radii.  This situation means that one cannot describe the field profile as a single perturbative expansion in $(m \rho)$ at all radii.\footnote{This issue was originally noted \cite{Affleck:1980mp}, and further studied in \cite{Nielsen:1999vq}, where the failure of the naive perturbative approach was shown explicitly.}
However, our ultimate goal is to compute the action of the solution, which as we will see does admit an organised expansion in $m \rho$.

To set up this expansion {we follow \cite{Affleck:1980mp}} and split the  solution into two regimes: 
the \emph{core solution} $\phi_\mathrm{core}(r)$ which is valid when $r\ll m^{-1}$, and 
the \emph{tail solution} $\phi_\mathrm{tail}(r)$ which is valid when $r\gg \rho$.
When $m\rho\ll 1$, these two regimes overlap, and we can determine the integration constants of both solutions by matching them at an intermediate radius $L$ which satisfies $\rho\ll L \ll m^{-1}$.

The field is thus described in a piecewise manner
\begin{equation}
    \phi(r) = \begin{cases}
    \phi_\mathrm{core}(r) & r \leq L ~, \\
    \phi_\mathrm{tail}(r) & r > L~.
    \end{cases}
\end{equation}
At very small radii the field is expected to 
approach the Fubini solution \eqref{equ:fubini},
\begin{equation}
    \phi_\mathrm{core}(r) \sim \phi_0(r) = \frac{4 \sqrt 3 \rho \lambda^{-1/2}}{\rho^2 + r^2},~\text{as}~{r\ll \rho}~.
    \label{phi-core-solution}
\end{equation}
This is valid if the quartic term dominates in the equation of motion \eqref{eom-constrained}. The mass term is subdominant as a consequence of the assumption $m\rho\ll 1$,
and the Lagrange multiplier term is subdominant if
\begin{equation}
\label{equ:smallkappa}
    \kappa\ll\lambda^{(n-2)/2}\rho^{n-4}~,
\end{equation}
where $n = 3, 6$ is the exponent of the {monomial} constraint.
We will check that this expression is satisfied at the end of the section.
For the wider range of radii, $r\ll m^{-1}$, we will extend eq.~(\ref{phi-core-solution}) perturbatively as
\begin{equation}
\label{equ:phicorecorr}
    \phi_\text{core}(r)=\phi_0(r)+\delta\phi_\text{core}(r)~,
    ~r\ll m^{-1}~.
\end{equation}
We will determine the correction $\delta\phi_\text{core}(r)$ shortly, but it is useful to discuss the tail solution first.

At large radii, far past the point where the field has rolled to near zero ($r \gg \rho$), the mass term dominates and the equation of motion is approximately
\begin{equation}
    \phi'' + \frac{3}{r} \phi' = m^2 \phi~.
\end{equation}
The solution to this may be written in terms of Bessel functions.  Taking the solution that decays to zero as $r \to \infty$ we have
\begin{equation}\label{eq:phi-tail-bessel}
    \phi_\mathrm{tail}(r) = B \frac{K_1(m r)}{m r}~,
\end{equation}
where $K_1(m r)$ is a modified Bessel function of the second kind,
and $B$ is a constant of integration which is fixed by matching
onto the large-$r$ limit of eq.~\eqref{phi-core-solution}. 
The small-$r$ limit of eq.~\eqref{eq:phi-tail-bessel} reads
\begin{equation}
    \phi_\mathrm{tail}(r) \sim B \left( \frac{1}{(m r)^2} + \frac{1}{2} \ln C m r + O\left( (mr)^2\ln mr \right) \right), ~\text{as}~r\ll m^{-1}~.
    \label{phi-tail-expansion-1}
\end{equation}
where the constant $C$ is defined by $\ln C \equiv \gamma - \ln 2 - \frac{1}{2} $ and $\gamma$ is {the Euler-Mascheroni constant}.
Matching this to the large-radius limit $r \gg \rho$ of the core solution eq.~\eqref{phi-core-solution} we find {
\begin{equation}
    B = 4 \sqrt{3} \rho \lambda^{-1/2} m^2~.
\end{equation}
}

The quantity of interest from the perspective of tunnelling calculation is the action of the constrained solution, rather than the exact field profile. Therefore, next we compute the first few terms of the action and the constraint in the $(m \rho)$ expansion.  We begin by rearranging eq.~\eqref{phi-tail-expansion-1} to emphasize the $(m \rho)$ dependence:
\begin{equation}
    \phi_\mathrm{tail}(r) \sim
    \frac{4 \sqrt 3  \rho  }{\lambda^{1/2}r^2} + 
    \frac{2 \sqrt 3 (m \rho)^2}{\lambda^{1/2}} \frac{\ln m \rho}{\rho} + 
    \frac{2 \sqrt 3}{\lambda^{1/2}} (m \rho)^2 \frac{\ln (C r / \rho)}{\rho}~,~\text{as}~r\ll m^{-1}~.
    \label{phi-expanded-tail}
\end{equation}

The expansion eq.~\eqref{phi-expanded-tail} is valid for intermediate radii, $\rho\ll r\ll m^{-1}$. In this same range, we can also use the core solution eq.~\eqref{equ:phicorecorr} by extending the asymptotic result eq.~\eqref{phi-core-solution} perturbatively in powers of $m\rho$.
The insight developed in~\cite{Affleck:1980mp} is to assume
that the structure of that expansion is the same as in eq.~\eqref{phi-expanded-tail} so as to smoothly match on to the tail solution.  This implies a solution of the form
\begin{equation}
    \phi_\mathrm{core}(r) = \phi_0(r) + (m \rho)^2 \ln(m \rho) \phi_1(r) + (m \rho)^2 \phi_2(r)~,
    \label{phi-m-rho-expansion}
\end{equation}
where the $\phi_i(r)$ functions are independent of $(m \rho)$.
The asymptotic behavior of the $\phi_i$ at $r\gg\rho$ is determined by the requirement to match onto the tail solution at large radii,
\begin{equation}
    \phi_0 \sim \frac{4 \sqrt 3 \rho }{\lambda^{1/2} r^2}~, 
    \quad \phi_1 \sim \frac{2 \sqrt 3 }{ \lambda^{1/2} \rho}~, \quad \phi_2 \sim \frac{2 \sqrt 3 }{\lambda^{1/2}\rho} \ln(C r / \rho)~,
    ~\text{as}~r\gg\rho~.
    \label{phi-perturbation-mrho}
\end{equation}
We also expect to recover the instanton solution in the massless theory as $r \to 0$,
\begin{equation}
    \phi_0 \sim \frac{4 \sqrt{3} \rho \lambda^{-1/2}}{\rho^2 + r^2}~, \quad \phi_1 \to 0~, \quad \phi_2 \to 0~,
    ~\text{as}~r\ll\rho~.
    \label{phi-perturbation-mrho-small}
\end{equation}
{It should be noted that explicit forms for $\phi_{1,2}$ were recently obtained for the hexic constraint~\cite{Aoki:2026zif}, while here we only require the asymptotic forms.  These asymptotic forms are sufficient for computing the leading-order corrections to the action.}

\begin{figure}[t]
    \centering
    \includegraphics[width=0.75\linewidth]{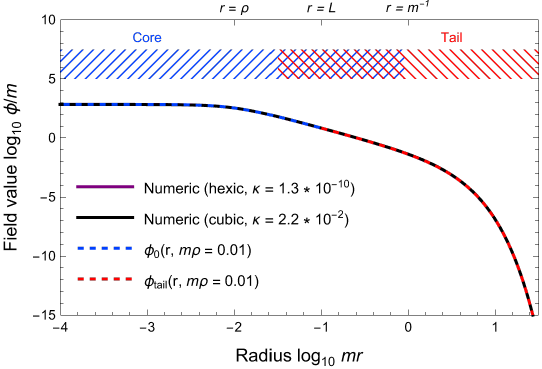}
    \caption{\small Comparison between  {numerical} and analytical solutions for a small instanton, in the particular case $\lambda = 1$.  The numerical solutions, for both the hexic and cubic constraints, are virtually indistinguishable.  The $\kappa$ in each case are selected so as to give an instanton size of $m \rho \approx 0.01$.  In the core region, the Fubini instanton~\eqref{equ:fubini} is used, and in the tail region, $\phi_\mathrm{tail}$ given by eq.~\eqref{eq:phi-tail-bessel} is used.  Good agreement between the solutions is observed in the appropriate limits.} 
    \label{fig:small-instanton-comparison}
\end{figure}

We have thus found the approximate field profiles $\phi_\mathrm{core}$ and $\phi_\mathrm{tail}$ given by eqs.~\eqref{phi-m-rho-expansion} and \eqref{eq:phi-tail-bessel}, respectively.  This is plotted in figure~\ref{fig:small-instanton-comparison}, for a particular instanton size of $\rho \approx 10^{-2} m^{-1}$. {Unlike what was presented in \cite{Elder:2025kya},} this is not a fit -- rather, the value of $\rho$ is found by evaluating the above equation at $r = 0$ to obtain
\begin{equation}
    \rho = \frac{4 \sqrt{3}}{\phi(0)}~,
    \label{numerical-instanton-size}
\end{equation}
where $\phi(0)$ is supplied by the numerical solution. 
The core solution depicted in the figure uses only the Fubini instanton eq.~\eqref{equ:fubini}, and not the corrections of the latter two terms in eq.~\eqref{phi-m-rho-expansion}.  Even without those corrections, there is excellent agreement with the numerical solutions, in both core and tail regions.

Having found the leading order dependence of the field on $(m \rho)$ we can now compute the action.  We'll begin by splitting the action into two pieces, corresponding to the contributions from the core area $0 < r < L$ and the tail area $L < r < \infty$,
\begin{align}
    S = S_\mathrm{core} + S_\mathrm{tail}~.
\end{align}
We compute $S_\mathrm{core}$ first,
\begin{equation}
    S_\mathrm{core} = \int_{r<L} d^4 x \left(\frac{1}{2} (\partial \phi_\mathrm{core})^2 + \frac{1}{2} m^2 \phi_\mathrm{core}^2 - \frac{\lambda}{4!}\phi_\mathrm{core}^4\right)~.
\end{equation}
To evaluate this, we
work in the $L \gg \rho$ limit.
The integral is dominated by the original terms and the Fubini instanton:
\begin{align} \nonumber
    S_\mathrm{core} &= \int_{r<L} d^4 x \left(\frac{1}{2} (\partial \phi_0)^2 - \frac{\lambda}{4!} \phi_0^4 \right) + \delta S~, \\ 
    &= \frac{16 \pi^2}{\lambda} - \frac{96 \pi^2}{\lambda} \left(\frac{\rho}{L}\right)^2
    + O\left(\frac{\rho}{L}\right)^6 + \delta S~.
    \label{S-core}
\end{align}
The second term is a correction due to the truncation of the integral at $r = L$, and is important to include at this stage because it will cancel with other terms later. There are three distinct contributions to the final term,
\begin{equation}
    \delta S=\delta S_0+\delta S_1 + \delta S_2~,
\end{equation}
which we consider in turn.

First there is a correction from the mass term, which at lowest order in $(m \rho)$ is
\begin{align} \nonumber
    \delta S_0 &= \int_{r<L} d^4 x \frac{1}{2} m^2 \phi_\text{core}^2~, \\ \nonumber
    &\approx
    \int_{r<L} d^4 x \frac{1}{2} m^2 \phi_0^2~,\\ 
    &\approx \frac{48 \pi^2}{\lambda} (m \rho)^2 \left(\ln m L - \ln m \rho - \frac{1}{2}\right)~.
\end{align}

The remaining terms come from 
the perturbation $\delta\phi_\text{core}(r)$ defined in eq.~(\ref{equ:phicorecorr}),
\begin{align} \nonumber
\delta S_1+\delta S_2
&= \int_{r<L} d^4x \left( \partial \phi_0 \partial \delta \phi_\text{{core}} - \frac{\lambda}{3!} \phi_0^3 \delta \phi_\text{{core}} \right)~, \\ \nonumber
    &= \int_{r<L} d^4 x \left( (-\Box \phi_0 - \frac{\lambda}{3!} \phi_0^3) \delta \phi_\text{{core}} + \partial(\delta \phi_\text{{core}} \partial \phi_0) \right)~, \\
    &= 2 \pi^2 L^3 \delta \phi_\mathrm{core}(L) \phi_0'(L)~,
\end{align}
where we have integrated by parts to arrive at the second line.
In the second line, the first term vanishes by the $\phi_0$ equation of motion~\eqref{fubini-eom}, and the second term is a total derivative, which does not vanish because $\delta \phi_\mathrm{core}$ remains finite at the boundary $r = L$.  Evaluating this surface integral at the large but finite radius $\rho \ll L \ll m^{-1}$ gives the third line.

Given eq.~\eqref{phi-m-rho-expansion}, we see that $\delta \phi_\mathrm{core}$ is composed of two terms.  It is only required to evaluate these terms at $r = L$.  Using eq.~\eqref{phi-perturbation-mrho}, we have
\begin{equation}
    \delta \phi_\mathrm{core}(L) \equiv (m \rho)^2 \ln (m \rho) \phi_1(L) + (m \rho)^2 \phi_2(L)~,
\end{equation}
We can now evaluate the perturbations to the action:
\begin{align} \nonumber
    \delta S_1 &= 2 \pi^2 L^3 (m \rho)^2 \ln (m \rho) \phi_1(L) \phi'_0(L)~, \\
    &= - \frac{96 \pi^2}{\lambda} (m \rho)^2 \ln m \rho~,
\end{align}
where we have used eq.~\eqref{phi-perturbation-mrho} in the second line.
Similarly,
\begin{align} \nonumber
    \delta S_2 &= 2 \pi^2 L^3 (m \rho)^2 \phi_2(L) \phi'_0(L)~,\\
    &= - \frac{96 \pi^2}{\lambda} (m \rho)^2 \left(\ln m L - \ln m \rho + \ln C\right)~.
\end{align}
Adding all of the $\delta S_i$ together we have
\begin{equation}
    S_\mathrm{core} = \frac{16 \pi^2}{\lambda} - \frac{96 \pi^2}{\lambda} \left( \frac{\rho}{L} \right)^2
    - \frac{48 \pi^2}{\lambda} (m \rho)^2 \left( \ln m \rho + \ln m L  + 2 \gamma - \ln 4 - \frac{1}{2} \right)~.
\end{equation}
In the expression we have used the definition of the constant $C$.
This differs from the solution in~\cite{Affleck:1980mp},
where a cancellation $\delta S_0 + \delta S_2 = 0$ was assumed. We have found that this assumption does not hold.

We have thus found a factor of 2 difference in the $(m \rho)$-dependent term, and an extra term proportional to $\ln mL$.  This term presents some difficulties, because (apart from appearing quite artificial) $L$ is related to $\rho$ such that this term diverges in the $m \rho \to 0$ limit because of our assumption that $mL \ll m\rho$.  The resolution is that one must also account for the contribution to the action in the tail region,
\begin{equation}
    S_\mathrm{tail} = \int_{r>L} d^4 x \left(\frac{1}{2} (\partial \phi)^2 + \frac{1}{2} m^2 \phi^2 - \frac{\lambda}{4!} \phi^4\right)~.
\end{equation}
This is to be evaluated using $\phi_\mathrm{tail}$ from eq.~\eqref{eq:phi-tail-bessel}. Since $\phi_\mathrm{tail} \sim r^{-2}$ when $r \ll m^{-1}$, we see that the mass term leads to a logarithm.
We shall see shortly that this logarithm precisely cancels the problematic term in $S_\mathrm{core}$.  

We first integrate the kinetic term by parts, picking up a surface term:
\begin{equation}
    S_\mathrm{tail} = \int_{r>L} d^4 x \left( - \frac{1}{2}\phi_\mathrm{tail} \partial^2 \phi_\mathrm{tail} + \frac{1}{2} m^2 \phi_\mathrm{tail}^2 - \frac{\lambda}{4!} \phi_\mathrm{tail}^4  + \frac{1}{2} \partial(\phi_\mathrm{tail} \partial \phi_\mathrm{tail}) \right)~.
\end{equation}
The first and second terms cancel by the $\phi_\mathrm{tail}$ equation of motion.  It can be seen that the quartic term is proportional to $(m \rho)^4$ by direct substitution of $\phi_\mathrm{tail}$ and therefore is higher order than we are dealing with presently:
\begin{equation}
    \int 2 \pi^2 r^3 dr \frac{\lambda}{4!} \phi^4_\mathrm{tail} \sim \frac{(m \rho)^4}{\lambda} \int_{m L}^{\infty} \frac{K_1(u)^4}{u} du~.
\end{equation}
That leaves the boundary term, which is a total derivative:
\begin{align} \nonumber
    S_\mathrm{tail} &= 
    \frac{1}{2} \int_{r>L} \partial (\phi_\mathrm{tail} \partial \phi_\mathrm{tail})~, \\
    &= \lim_{r \to \infty} \pi^2 r^3 \phi_\mathrm{tail}(r) \phi_\mathrm{tail}'(r) - \pi^2 L^3 \phi_\mathrm{tail}(L) \phi_\mathrm{tail}'(L)~.
\end{align}
{The first term goes to $0$ as $r \to \infty$}, and after using $m L \ll 1$ we have
\begin{equation}
    S_\mathrm{tail} \approx \frac{96 \pi^2}{\lambda} \left(\frac{\rho}{L} \right)^2 +  \frac{48 \pi^2}{\lambda} (m \rho)^2 \left(\ln m L + \gamma - \ln 2 - 1 \right)~.
\end{equation}
The first two terms precisely cancel the first two corrections in $S_\mathrm{core}$.
Adding up the contributions, we are left with 
\begin{equation}
    S \approx \frac{16 \pi^2}{\lambda} - \frac{48 \pi^2}{\lambda} (m \rho)^2 \left(\ln m \rho + \gamma - \ln 2 + \frac{1}{2} \right)~.
    \label{action-small-instanton}
\end{equation}
The correction to the action is independent of the constraint used, and the Lagrange multiplier, because the constraint term is subdominant to the quartic term in the potential.

\begin{figure}
    \centering
    \includegraphics[width=0.5\linewidth]{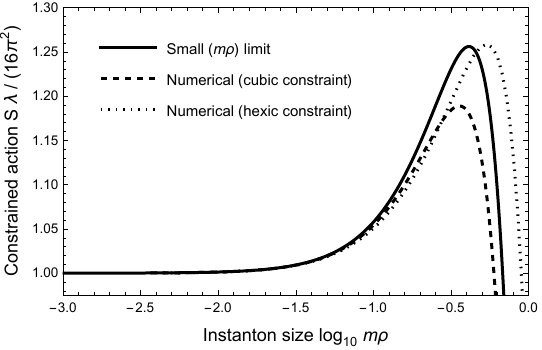}
    \caption{\small Comparison between the small-instanton limit of eq.~\eqref{action-small-instanton} and the numerical results from a shooting method for both types of constraint used in this work.  Excellent agreement is found in the small-instanton limit, where the instanton size $\rho \ll m^{-1}$.}
    \label{fig:small-instanton-action-vs-size}
\end{figure}

We compare to numerical results for action using the cubic and hexic constraints in figure~\ref{fig:small-instanton-action-vs-size}.  As one would expect excellent agreement is seen in the small instanton limit $\rho \ll m^{-1}$. Note that, while the size of a constrained solution is generally ambiguous, it makes sense in this particular limit and for these particular constraints, with the size defined via eq.~\eqref{numerical-instanton-size}.  We also note that this expression for the action agrees precisely with that obtained via the continuous field profiles derived for the hexic constraint in \cite{Aoki:2026zif}.

\begin{figure}[t]
    \centering
    \includegraphics[width=0.475\linewidth]{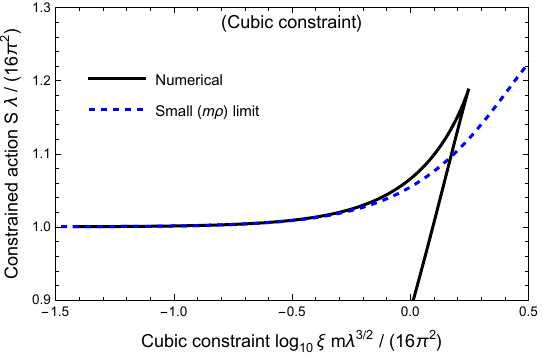}
    \includegraphics[width=0.475\linewidth]{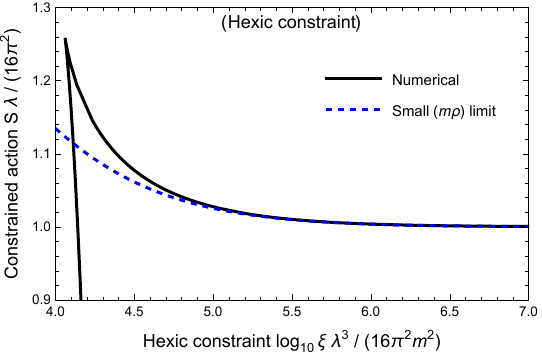}
    \caption{Action vs. constraint, for both types of constraint used in this work.}
    \label{fig:small-instanton-action-vs-constraint}
\end{figure}

Finally, we may evaluate the constraints using eq.~\eqref{constraint-definition}.
The constraints ${\cal O}$ were chosen to go to $0$ sufficiently quickly that the integral is finite, so we can evaluate these at leading order by simply using $\phi \approx \phi_0$ and taking $L \to \infty$:
\begin{align} \nonumber
    \bar \xi_3[\phi] &\approx 96 \sqrt{3} \pi^2 \frac{(m \rho)}{m\lambda^{3/2}}~, \\
    \bar \xi_6[\phi] &\approx \frac{27648 \pi^2 m^2}{5 (m \rho)^2 \lambda^6}~.
    \label{constraint-values}
\end{align}
We have plotted the action against the constraint in figure~\ref{fig:small-instanton-action-vs-constraint}.  Once again good agreement is found with the numerical solutions of \cite{Elder:2025kya} within the regime of validity of the small-$m\rho$ expansion.

We are now in a position to check our original claim, that the constraint term in the modified action could be treated perturbatively.  From eq.~\eqref{action-general-constraint} we know that $\kappa$ must satisfy
\begin{equation}
    \kappa = - \frac{\partial S}{\partial \bar \xi}~.
\end{equation}
We use eq.~\eqref{constraint-values} to obtain $(m \rho)$ in terms of $\bar \xi$, substitute into the action eq.~\eqref{action-small-instanton}, and then differentiate with respect to $\bar \xi$.  We find, for each type of constraint,
\begin{align} \nonumber
    \kappa_3 &\approx m \sqrt \frac{\lambda}{3} (m \rho) \log m \rho~, \\
    \kappa_6 &\approx - \frac{\lambda^6}{576 m^2} (m \rho)^4 \log m \rho~.
    \label{kappa-explicit}
\end{align}
We see that eq.~\eqref{equ:smallkappa} is easily satisfied, justifying our perturbative treatment of the constraint.

\section{Large $(m \rho)$ limit: $\phi^6$ constraint and the thin-wall approximation}
\label{sec:large-hexic}

If the modified potential \eqref{modified-potential} has two nearly-degenerate minima, separated by a large barrier,
then constrained solutions will be very large. That is, the transition between the false and true vacua of the modified potential occurs at a very large but finite radius. In this case we can apply the classic thin-wall approximation of instantons, as done in \cite{Coleman:1977py}, although for a different potential so we recount the main points here.

Intuitively this approach may be understood using an analogy to the motion of a ball in classical mechanics.  In this analogy, the equation of motion \eqref{eom-constrained} is interpreted as that of a ball in a 1D potential, where the field value $\phi$ is the position of the ball and radius $r$ plays the role of time.  Then the potential energy of the ball is given by the negative  modified potential $-\tilde V_\kappa(\phi)$, and the $\phi'/r$ term acts as a drag term that decays away with time. The ball starts out near the true vacuum, near the top of a hill in the (inverted) potential (see figure~\ref{fig:potentials} for an illustration of the shape).  It must end up on top of the hill corresponding to the false vacuum as $r \to \infty$, which in this case is at nearly the same height as the starting point. It stands to reason that the transition must happen at a very large time (i.e. radius), if it is to reach its endpoint, otherwise it will lose too much energy to the drag term.  We can therefore neglect the drag term in the equation of motion for the region around the transition between the false and true vacua. Before the transition happens, the field rolls very slowly and remains near its starting point. After the transition completes, it approaches the false vacuum asymptotically. Thus, combining the solutions in the three regimes, the field profile can be described in its entirety.

This technique will be demonstrated for the $\phi^6$ constraint.  It will not work for the $\phi^3$ constraint due to the lack of two near-degenerate minima, so we handle that case via a different method  in the next section. 
We start by splitting the constrained potential into two pieces: one with degenerate minima and a small perturbation via
\begin{equation}
    \tilde V_\kappa(\phi) \equiv V_+(\phi) + \delta V(\phi)~,
\end{equation}
where
\begin{equation}
    V_+(\phi) = \frac{1}{2} m^2 \phi^2 - \frac{\lambda}{4!} \phi^4 + \kappa_\mathrm{max} \phi^6~.
\end{equation}
We have defined $\kappa_\mathrm{max}$ as the value of $\kappa$ at which the local minima are degenerate,
which corresponds to $\kappa_\mathrm{max} =
\lambda^2/(1152m^2)$ and the ``symmetric potential'' has degenerate minima at $\phi_\pm = \pm 2 \sqrt 6 m \lambda^{-1/2}$.  Consequently, $\kappa \to \kappa_\mathrm{max}$ corresponds to the large-$(m \rho)$ limit.
Above $\kappa_\mathrm{max}$ the local minimum at $\phi = 0$ becomes the true vacuum, so this represents the maximum value of $\kappa$ that need be considered for our purposes.  We expand in this limit via $\kappa = \kappa_\mathrm{max} - \epsilon$, where $\epsilon\ll\kappa_{\rm max}$, so the potential perturbation is
\begin{equation}
    \delta V(\phi) = -\epsilon \phi^6~.
\end{equation}

Since the constrained solutions are large, we can neglect the $\phi'$ term in the equation of motion, and we will also neglect the perturbation to the symmetric potential, so the equation of motion is
\begin{equation}
    \phi''(r) = \frac{\rm d V_+(\phi)}{{\rm d} \phi}~.
\end{equation}
This equation of motion gives the ``kink solution'' which smoothly interpolates between the minima in the limit that the $\phi'$ drag term is negligible.  The method of solution is well-known (see~\cite{Coleman:1977py, Goldstone:1974gf, Christ:1975wt}) which we outline here.

We are interested in solutions that satisfy the boundary conditions $\phi(0) = \phi_+$ and $\phi(\infty) = 0$ (we could equivalently have chosen $\phi(0) = \phi_-$).
The above equation of motion may be integrated once to obtain
\begin{equation}
    \frac{1}{2}\phi'^2 - V_+(\phi) = \mathrm{const}~.
\end{equation}
The field starts at $r = 0$ with $\phi' = V_+(\phi_+) = 0$ so the integration constant is zero.  Rearranging, we find
\begin{equation}
    \frac{d \phi}{dr} = - \sqrt{2 V_+(\phi)}~,
    \label{kink-eom-integrated-once}
\end{equation}
where we have chosen the negative root because the field rolls from large to small $\phi$ as $r$ increases.

Qualitatively, the field sits with nearly zero gradient at $\phi = \phi_+$ from $r = 0$ to $r = R$, where $R$ is the approximate size of the constrained solution.\footnote{Although $R$ is similar in spirit to the parameter $\rho$ of the previous sections, they are not mathematically equivalent so we use a distinct label.}  It then quickly rolls to $\phi = 0$ and remains there out to $r = \infty$.  Integrating eq.~\eqref{kink-eom-integrated-once} one more time, we find the kink solution $\phi_\mathrm{kink}$  to be
\begin{equation}
    r = \int_{\phi_\mathrm{kink}(r)}^{\phi_+} \frac{d \phi}{\sqrt{2 V_+(\phi)}}~.
\end{equation}
Note that we have slightly adjusted the sign and limits of the integral relative to~\cite{Coleman:1977py} in order to match our specific setup.
The overall strategy remains the same, however.
The kink action may now be computed as
\begin{align} \nonumber
    S_\mathrm{kink} &= \int_0^\infty dr \left( \frac{1}{2}\phi'^2 + V_+(\phi) \right)~,\\
    &= \int_{0}^{\phi_+} d\phi \sqrt{2 V_+(\phi)}~.
\end{align}
For the $\phi^6$ constraint at hand we have the very simple result
\begin{equation}
    S_\mathrm{kink} = \frac{6 m^3}{\lambda}~.
\end{equation}

\begin{figure}[t]
    \centering
    \includegraphics[width=0.75\linewidth]{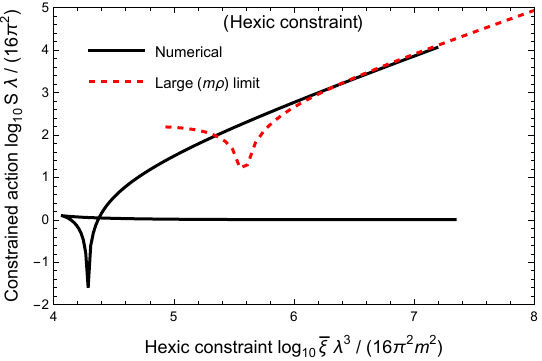}
    \caption{\small Comparison of the thin wall approximation to numerical results.  The upper branch is the large instanton limit, and excellent agreement is found in that regime.  The red line is plotted for $0.6 < \kappa / \kappa_\mathrm{max} < 0.99$, which corresponds to instanton sizes of $3 \lesssim m \rho \lesssim 10^3$.}
    \label{fig:action-vs-constraint-large-hexic}
\end{figure}

The action of the constrained solution may be computed as follows.  The full action is split into two pieces, the bulk region between $r = 0$ to $R$ where $\phi' \approx 0$ as well as boundary where the field suddenly rolls $\phi \to 0$ at $r \approx R$.\footnote{The contribution from region $r \gg R$ vanishes trivially.} These are well-approximated by a constant field and the kink solution, respectively:
\begin{align} \nonumber
    \tilde S_\kappa &= \int_0^\infty 2 \pi^2 r^3 dr \left( \frac{1}{2} \phi'^2 + \tilde V_\kappa(\phi) \right)~, \\
    &= \frac{\pi^2}{2} R^4 \delta V(\phi_+) + 2 \pi^2 R^3 S_\mathrm{kink}~.
\end{align}
The first term gives the contribution from the bulk, due to the field being essentially constant at the false vacuum.  The second term is proportional to the surface area of the 4-dimensional ball of radius $r=R$.  
Extremising this action with respect to the instanton size $R$ we find
\begin{equation}
    R = - \frac{3 S_\mathrm{kink}}{\delta V(\phi)} = \frac{3 S_\mathrm{kink}}{\epsilon \phi_+^6}~.
\end{equation}
As we expected as $\epsilon \to 0$ the instanton becomes arbitrarily large.
The action may now be written as
\begin{equation}
    \tilde S_\kappa \approx - \frac{27 \pi^2 S_\mathrm{kink}^4}{2 \delta V^3} \sim + \epsilon^{-3}~.
\end{equation}
Note that the action we have computed at this point is the action in the theory defined by the modified potential. However, in the constrained instanton framework, the action we are interested in is the action without the $\kappa$ term. Therefore, to obtain the constrained solution action, we need to subtract the contribution from the constraint term.
That is, from eq.~\eqref{action-general-constraint} we have
\begin{equation}
    S = \tilde S_\kappa - \kappa \bar \xi~.
\end{equation}
Therefore, to compute the action, we need to compute the constraint term.  At leading order we may use $\kappa = \kappa_\mathrm{max}$ and truncate the integral at $r = R$ to give
\begin{equation}
    \bar \xi \approx \frac{\pi^2}{2} R^4 \phi_+^6 \sim \epsilon^{-4}~.
\end{equation}
Based on the scaling with $\epsilon$, it is clear that as $\epsilon \to 0$ the second term in $S$ becomes the dominant one.  The full action is plotted in figure~\ref{fig:action-vs-constraint-large-hexic}, where excellent agreement with the numerical solutions is observed in the limit of large instanton size.

\section{The $\phi^3$ constraint and the large $\kappa$ limit}
\label{sec:large-cubic}

In the previous examples we had instanton size as a clear parameter driving our approximations.  In the case of a large constrained solution, we were able to use the classic thin-wall approximation because the false and true vacua were nearly degenerate.  This translated into the possibility of a transition between vacua at very large radii.   In the case of the $\phi^3$ constraint, however, there remains only one local minimum, so clearly another approach is required.  In the intuitive analogy to a ball in a 1D potential, the ball starts at rest some way up the hill at a finite value of $\phi$. (The reader may wish to refer to figure~\ref{fig:potentials} as a visualisation aid.)  Because the slope of the modified potential is large at the starting point, the ball starts to roll immediately.  It loses a great deal of its potential energy to drag, eventually arriving at $\phi = 0$ as $r \to \infty$.  There is thus no clear definition of instanton size in this case: the ball starts rolling immediately, regardless of the size of the constraint term.
What we will find is that even in the limit of arbitrarily large $\kappa$ the overall shape of the instanton field profile does not vary dramatically.

This is therefore a good demonstration of a case where the size is not well-defined, i.e. there is not a clear $(m\rho) \gg 1$ limit.  Rather, the full-width at half-maximum of the field profile {(one of the possible measures of the constrained solution size)} remains $\mathcal{O}(1)$ throughout the regime of interest, provided one uses properly scaled coordinates which will be introduced shortly.  We will see that in the the large $\kappa$ {limit} the quartic interaction becomes negligible.  While not as intuitive as the previous results, with less clear connections to quantities like instanton size, it is nevertheless possible obtain precise results in this limit. 

\begin{figure}
    \centering
    \includegraphics[width=0.75\linewidth]{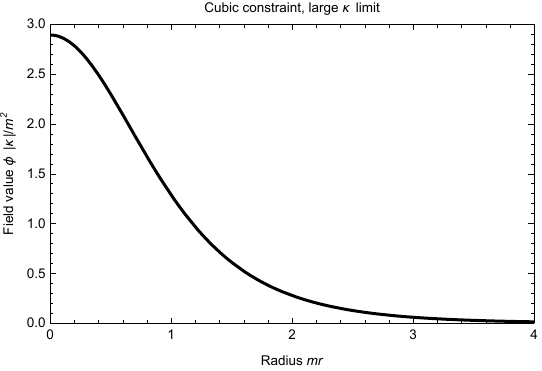}
    \caption{Field profile for a constrained solution in the limit of large $\kappa$, which solves eq.~\eqref{eom-large-cubic}.}
    \label{fig:large-cubic}
\end{figure}

We will extract the dependence of the action and constraint on the parameters in the limit of a large Lagrange multiplier {$\kappa$}.  The modified action is, for the cubic constraint,
\begin{equation}
    \tilde S_\kappa = \int d^4 x \left( \frac{1}{2}(\partial \phi)^2 + \frac{1}{2} m^2 \phi^2 + \kappa \phi^3 - \frac{\lambda}{4!} \phi^4 \right)~.
\end{equation}
The cubic term breaks the $\phi \to - \phi$ symmetry, so we will assume $\kappa < 0$ without loss of generality.  This means that the most probable escape path out of the $\phi = 0$ false vacuum is to positive $\phi$.

We wish to evaluate this in the limit of large $|\kappa|$.  We will see that this corresponds to small $\phi$ everywhere.  To make use of these observations, we can rescale the action into dimensionless variables
\begin{equation}
    \hat x_i = m x_i~, \quad \quad \hat \phi = \frac{|\kappa|}{m^2} \phi~.
\end{equation}
Under this rescaling, the modified action becomes
\begin{equation}
   \tilde S_\kappa = \frac{m^2}{\kappa^2} \int d^4 \hat x \left( \frac{1}{2} (\hat \partial \hat \phi)^2 + \frac{1}{2} \hat \phi^2 - \hat \phi^3 - \frac{\lambda}{4!} \frac{m^2}{\kappa^2} \hat \phi^4 \right)~,
\end{equation}
where $\hat \partial_i$ indicates a partial derivative with respect to the dimensionless coordinates $\hat x^i$.
This makes it clear that at large $|\kappa|$, the field $\hat \phi$ is ${\cal O}(1)$ and we are justified in dropping the quartic term if $\lambda \ll m^2 / \kappa^2$.  More precisely, we require the quartic term to be smaller than other terms in the potential, which means we need
$\hat \phi \ll \kappa^2 / (m^2 \lambda)$ and $\hat \phi \ll \kappa / (m \sqrt \lambda)$, whichever is more stringent.  Since we expect that this is the limit of large $\kappa$ and small $m$ and/or small $\lambda$, it is likely the latter inequality that is more difficult to satisfy.

Under these assumptions, we are free to drop the quartic term entirely.  The modified action is then
\begin{equation}
    \tilde S_\kappa = \frac{m^2}{\kappa^2} \int d^4 \hat x \left( \frac{1}{2} (\hat \partial \hat \phi)^2 + \frac{1}{2} \hat \phi^2 - \hat \phi^3\right)~.
\end{equation}
This holds for all $\kappa, \lambda, m$ that satisfy the above assumptions.  In this limit there is only one bounce solution, and we evaluate the above integral for that solution numerically.   Note that unlike in section~\ref{sec:large-hexic} the $\phi'$ drag term cannot be neglected.  The bounce solution is found by solving the equation of motion for the modified action:
\begin{equation}
    \hat \phi'' + \frac{3}{\hat r} \hat\phi' = \hat \phi - 3 \hat \phi^2~.
    \label{eom-large-cubic}
\end{equation}
This is subject to the boundary conditions $\hat \phi'(0) = \hat \phi(\infty) = 0$.

\begin{figure}
    \centering
    \includegraphics[width=0.75\linewidth]{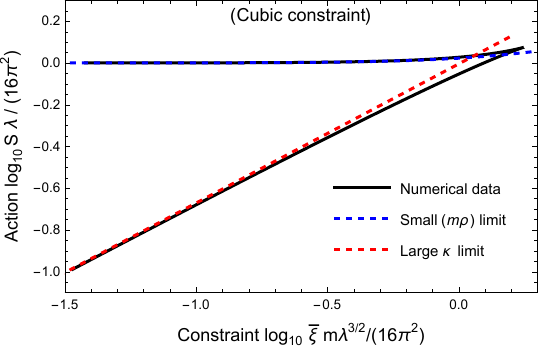}
    \caption{Comparison between the asymptotic limits of constrained solutions for the modified action with a cubic constraint.  The approximations are given by Eqs.~\eqref{action-small-instanton} and~\eqref{eq:cubic-constraint-large-kappa-results}, and are compared to the numerical results of~\cite{Elder:2025kya}.}
    \label{fig:cubic-S-vs-xi}
\end{figure}

This problem is solved with a shooting method~\cite{Press:2007ipz}, which guesses an initial value for $\hat \phi$ at the origin and integrates outwards.  A tolerance was chosen of $\delta_\mathrm{tol} = 10^{-10}$, and was solved on the finite range $\hat r \in [\delta_\mathrm{tol}, 20]$.  The initial condition $\hat \phi_0$ was bracketed by two values $\hat \phi_{0, \rm high}, \hat \phi_{0, \rm low}$, and iterated until $\hat \phi_{0, \rm high} - \hat \phi_{0, \rm low} < \delta_\mathrm{tol}$.  The boundary condition satisfaction was found to be $\hat \phi(\hat r_\mathrm{max}) = 5.2 \times 10^{-6}$~.  The numerical solution for $\hat \phi$ is shown in figure~\ref{fig:large-cubic}.  It is characterised by an initial value $\hat \phi_0 = 2.89$, and a half-width half-maximum of $\hat r_\mathrm{hwhm} = 0.914$.  Once the numerical solution is obtained, the action and constraint are computed from the above definitions, finding
\begin{equation} 
    S = 68.1 \frac{m^2}{\kappa^2}~, \quad \quad 
    \bar \xi = -45.4 \frac{m^2}{|\kappa|^3}~.
    \label{eq:cubic-constraint-large-kappa-results}
\end{equation}
This represents a complete solution of the problem in the limit of large Lagrange multiplier $|\kappa|$.
In figure~\ref{fig:cubic-S-vs-xi} we see that there is excellent agreement between these results and the numerical solutions of~\cite{Elder:2025kya}, within the appropriate limit.

\section{Discussion}
\label{sec:discussion}

In this work we have demonstrated three methods of analytic construction for constrained solutions, in three different limits, for two different types of constraint.  When the constrained solutions are small, they closely resemble instantons in the massless theory. We have followed the overall logic of~\cite{Affleck:1980mp}, with some corrections to the derivation which are verified via comparison to numerical solutions.  For the hexic $\phi^6$ constraint we find that in the limit where $\kappa$ is large we are able to apply the thin-wall approximation of \cite{Coleman:1977py} to describe large constrained solutions.  For the cubic constraint, we find that this same strategy does not work in the same limit.
Instead, we developed a scaling argument to simplify the problem and showed that one can use a single numerical solution to find all constrained solutions in this limit.

The reader is advised that, just as the absence of instanton solutions does not imply the absence of vacuum tunnelling, the existence of constrained solutions does not imply a non-negligible tunnelling rate.  Once a solution is obtained one must then calculate its contribution to the path integral. 
A given constrained solution may ultimately contribute very little (or nothing at all) to the path integral. This was explored in~\cite{Elder:2025kya} and will be further studied in future work \cite{prefactorPaper}.

{\bf Acknowledgements} 

B.E.\ and A.R. acknowledge support from STFC Consolidated grants ST/T000791/1 and ST/X000575/1.
B.E.\ was also supported by Simons Investigator award 690508, K.G. was supported by the STFC DTP research studentship grant ST/X508433/1,
and A.R.\ was supported by a CERN Scientific Associate position and a Visiting Fellowship at Wadham College, Oxford.

\bibliographystyle{JHEP}


\providecommand{\href}[2]{#2}\begingroup\raggedright\endgroup

\end{document}